\documentclass[a4paper,12pt]{article}
\usepackage[utf8]{inputenc}
\usepackage{graphicx}
\usepackage{hyperref}
\usepackage{amssymb}
\usepackage{amsfonts}
\usepackage{color}
\usepackage{amsmath}
\usepackage{tikz}

\def\sgn{\hbox{sgn}}
\newtheorem{theorem}{Theorem}
\newtheorem{corollary}{Corollary}

\begin{document}
\title{Stochastic properties of an inverted pendulum on a wheel on a soft surface.}
\author{O.M.~Kiselev}
\maketitle
\begin{abstract}
We study dynamics of the inverted pendulum on the wheel on a soft surface and under a proportional-integral-derivative controller. The behaviour of such pendulum is modelled by a system with a differential inclusion. If the the system has a sensor for the rotational velocity of the pendulum, the tilt sensor and the encoder for the wheel then this system is observable. The using of the observed data for the controller brings stochastic perturbations into the system.  The  properties of the differential inclusion under stochastic control is studied  for upper position of the pendulum.  The formula for the time, which  the pendulum spends near the upper position, is derived.
\end{abstract}

\section{Introduction}
\label{secIntroduction}

The wheeled inverted pendulum  (WIP) is a popular model for studies nor only  dynamics and the system  of the control for robotics equipments near instability positions. A list of contemporary works in this field is too large. Here we should mention studies for the derivation of the mathematical model for the WIP and the control synthesis, which one can see for example in the book \cite{Formalskii2016Eng}, and the article  \cite{MartynenkoFormalskii2013Eng}. The questions concerned an stability and control for WIP with two wheels were considered in  \cite{PathakFranchAgrawal2005} for the horizontal and in  \cite{SamiMichalskaAngeles2007} for the inclined surface.

\begin{figure}[h]
\begin{center}
\begin{tikzpicture}[ultra thick]

\draw (-1,0)--(-0.35,0.1);
\draw (0.55,0.2)--(1.5,0.4);
\draw[thin] (1.5,0.4)--(2.5,0.6);
\draw[thin] [->] (2.4,0) arc[start angle=0, end angle=13, radius=2.4];
\draw (2.6,0.3) node {$z$} ;

\foreach \i in {-1,-0.7,-0.4,-0.1,0.2,0.5,0.8,1.1}
	\draw (\i,\i*0.13+0.15)--(\i+0.3,\i*0.13+0.15-0.3);

\draw[fill=white] (0,1) circle [radius=1];
\draw(0,1)--(1,4);
\draw(0,1) circle [radius=0.1];
\draw[fill=white] (1,4) circle [radius=0.5];

\begin{scope}[thin]
\draw (0,0) -- (2.5,0);
\draw (2.1,-0.3) node {$x$} ;
\draw (0,0) --(0,5);
\draw (-0.2,5.1) node {$y$} ;
\draw[<-] (0,3) arc[start angle=90,end angle=77,radius=3];
\draw (0.3,3.3) node {$\alpha$} ;
\draw [->](1.3,1) arc[start angle=0,end angle=60,radius=1.3];
\draw (1.3,2) node{$\beta$};
\end{scope}
\end{tikzpicture}
\end{center}
\caption{The inverted pendulum on the wheel. Let us denote  $r$ is the radius of the wheel, $l$ is the pendulum length, $\alpha$ is the angle of the pendulum turn, $\beta$ is the angle of the wheel turn and $z$ is the inclination angle of the surface.}
\label{figurePendulumOnTheWheel}
\end{figure}
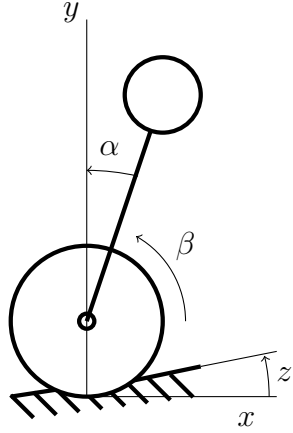

We apply the proportional-integral-derivative (PID) controller to stabilize the WIP at  the upper point.  We should mention the PID controller is often used to the objects of different nature \cite{AstromHagglund1994}. For example the PID controller can be used to stabilize the WIP on the hard horizontal, see \cite{HatakeyamaShimada2008},\cite{NasirRajaIsmailAhmad2010} and a review   \cite{ChanStolHalkyard2013}.  But the soft surface is more complicated to stabilize the WIP. The mathematical model for the WIP controlled by  the PID controller on the soft surface  was offered and detailed studied in \cite{Kiselev2020} (see also preprint \cite{Kiselev2020arxiv1}).

The major part of the PID controller is an observation value of the angle of the pendulum position. This angle can be found by gyroscope sensor . Usually the sensor obtains the value with a small stochastic error. The maximal amplitude and the dispersion for the error is  standardized by specification of the sensor, see for example  \cite{mpu6000}.

An additional sensor for the WIP is a tilt sensor. Such sensors are commonly used and have a detailed specification in which the interval of errors and the dispersion are pointed, see for example  \cite{MMA7361L}.

To obtain amount of the rotation of the wheel we use an encoder. If the wheel does not slip, the errors of the digital encoder appear due to round up only. Slipping brings additional errors to the value of the encoder. Therefore one of the problems for the controller is to detect the slipping.

In this work we show that the gyroscope and tilt for the pendulum and encoder for the wheel are enough to observe the state of the system in framework of the mathematical model.

However the noise of the sensors and the slipping bring stochastic errors into the value of the sensors. Therefore the controller which uses these values has stochastic perturbations. Hence the mathematical model with the digital PID controller is stochastic.

In section \ref{secEquationsOfMotion} we describe the mathematical model of WIP on the soft surface with the PID controller. The dynamic model on the soft surface contains a rolling resistance. Pure mathematically this resistance is described by  the differential inclusion,  
see \cite{Kiselev2020}.

In section \ref{secObservation} one can find the formulas for the current state of the WIP. These formulas use the data obtained from the gyroscope,  tilt sensor and encoder.

In section \ref{secStochasticPropOfObservedData} we discuss the stochastic properties of the data for the PID controller. The errors and dispersion of the data from the sensors are assumed as known form the technical notes.

In section \ref{secStochasticPropOfPendulum} we consider the properties of the WIP under the PID controller with stochastic perturbations. The perturbations appear in the control by using the data with the stochastic errors.

\section{Dynamical system for the WIP}
\label{secEquationsOfMotion}

Here we will consider the moving of the WIP with the additional control torque on the wheel. The torque is denoted by $u$. Let us assume that the equipment moves on soft surface with inclination $z$. The value of $z$ depends on the traversed path  by the wheel and hence one can write  $z=z(\beta)$. The mathematical model of such moving has the form (see \cite{Kiselev2020}, also preprint \cite{Kiselev2020arxiv1}):
\begin{eqnarray}
\ddot{\alpha}
&=&
 \sin(\alpha) -(\cos(\alpha-z)\ddot{\beta}+\sin(\alpha-z)\dot{\beta}^2)\rho 
-2\frac{\rho}{\zeta}u,
\nonumber\\
(\zeta +2)\rho\,\ddot{\beta}
& \in & 
F(\alpha,\dot{\alpha},\ddot{\alpha},\dot{\beta}).
\label{inclusionForPendulumOnWheelFullWithStabAndDump}
\end{eqnarray}
Let us denote
$$
f=-\sin(z)-\left(\ddot{\alpha}\cos(\alpha-z) -\dot{\alpha}^2\sin(\alpha-z)\right)  \zeta+\frac{2}{\rho}u.
$$ 
In formula  (\ref{inclusionForPendulumOnWheelFullWithStabAndDump}) the map  $F(\alpha,\dot{\alpha},\ddot{\alpha},\dot{\beta})$ has the form:
$$
F(\alpha,\dot{\alpha},\ddot{\alpha},\dot{\beta})
=
\left\{
\begin{array}{cc}
f-\nu\ \sgn(\dot{\beta}),& \quad\quad \{\forall\dot{\beta}\not=0\};
\\
(-\nu,\nu), & \quad\{\dot{\beta}=0\}\cup\{ |f|\le\nu\};
\\
f, & \quad \{\dot{\beta}=0\}\cup\{\{\alpha,\dot{\alpha},\ddot{\alpha}\} \in \{|f|>\nu\}\}. 
\end{array}
\right.
$$
Here the parameters of the mathematical model are following: $\alpha$  is an angle of the pendulum turn, $\beta$ is the angle of wheel turn, $z$ is the current inclination of the soft surface, $\nu$ is the torque of the friction resistant, $\rho=r/l$ is the ratio of the wheel radius and the length of the pendulum, $\zeta$ is the ratio of pendulum mass and the rim mass.

The system (\ref{inclusionForPendulumOnWheelFullWithStabAndDump}) can be simplify for hard  ($\nu=0$) surface with the constant inclination ($z\equiv\epsilon$). As a result one gets the second order equation for $\alpha$:
\begin{eqnarray}
(\sin^2(\epsilon-\alpha)\zeta +2)\ddot{\alpha}=
(2+\zeta)\sin(\alpha)  +
\sin(\epsilon)  \cos( \epsilon -\alpha) +
\nonumber\\
\frac{1}{2}\zeta\dot{\alpha}^2 \sin(2(\epsilon -\alpha))-
2\left(
\frac{1}{\rho}\cos(\epsilon-\alpha)+
\left(1+\frac{2}{\zeta}\right)\rho
\right) u. 
\label{EqPendulumOnWheel1WithStab}
\end{eqnarray}
The particular case for the WIP on the hard horizontal ($\epsilon=0$) looks as follow:\begin{eqnarray}
(\sin^2(\alpha)\zeta+2)\ddot{\alpha}
&=
&
(2+\zeta)\sin(\alpha)-\frac{1}{2}\zeta \dot{\alpha}^2\sin(2\alpha) -
\nonumber
\\
&\,&
2\left(\frac{1}{\rho}\cos(\alpha)+\left(1+\frac{2}{\zeta}\right)\rho\right)u
\label{EqPendulumOnWheel0WithStab}
\end{eqnarray}

The control torque with the PID controller has the following form:
$$
u=k_1\alpha+k_2\dot{\alpha}+k_3 A,\quad \text{where},\quad A\equiv\int^t \alpha(t) dt.
$$

In this case system  (\ref{inclusionForPendulumOnWheelFullWithStabAndDump})  has a particular solution:
\begin{eqnarray}
\alpha
\equiv0,
&
\quad 
\displaystyle
A=\frac{\sgn(\dot{\beta})\zeta\nu\rho}
{\left(2k_3\zeta+4k_3\right)\rho^2+2k_3\zeta},
\nonumber\\
\beta
=
\bigg\{
&
\begin{array}{c}\displaystyle
\beta_0+\beta_1(t-t_0)-\frac{\zeta\nu\rho\,\sgn(\dot{\beta})}
{\left(2k_3\zeta+4k_3\right)\rho^2+2k_3\zeta}
\frac{(t-t_0)^2}{2},\quad (t-t_0)<T;
\\
\displaystyle
\beta_0+\beta_1 T-
\frac{\zeta\nu\rho\,\sgn(\dot{\beta})}
{\left(2k_3\zeta+4k_3\right)\rho^2+2k_3\zeta}
\frac{T^2}{2},\quad (t-t_0)\ge T,
\end{array}
\label{partialSolution}
\end{eqnarray}
where
$$
T=\,\frac{1}{\beta_1}\,\frac{\zeta\nu\rho\,\sgn(\dot{\beta})}
{\left(2k_3\zeta+4k_3\right)\rho^2+2k_3\zeta},\quad 
\{t_0,\beta_0,\beta_1\}\in\mathbb{R}.
$$
There exists the set of the parameters  $\zeta,\rho,k_1,k_2,k_3$ when solution  (\ref{partialSolution}) is an attractor as  $(t-t_0)<T$ \cite{Kiselev2020} (see also preprint \cite{Kiselev2020arxiv1}).

In an ideal case the control should be defined by the current  values of $\alpha,\dot{\alpha},A$, but for real equipment these parameters can be obtained using the sensors at  the moment $t_i$, where $i\in\mathbb{N}$. As a result the control is a discrete function: $u(t)=u(t_i)=u_i$.

At the interval $t\in(t_i,t_{i+1})$  the control torque  $u_i$  is a constant. Such system has a first integral and can be integrate in quadratures. 

For example the moving on the hard surface with the constant inclination (\ref{EqPendulumOnWheel1WithStab}) has a first integral at the interval  $t\in(t_i,t_{i+1})$:

\begin{eqnarray}
\mathcal{E}_i=
&
\cos(\alpha)(\zeta +2)
+\sin(\epsilon)\sin(\epsilon-\alpha)+\left(\frac{1}{2}
\,\sin^2(\epsilon-\alpha)\zeta+1\right)\dot{\alpha}^2+
\nonumber\\
&
\left(2\left(1+\frac{2}{\zeta}\right)\rho\alpha -\frac{2}{\rho}\sin(\epsilon-\alpha))\right)u_i 
.
\label{formulaForConservationLawOnSlantedSurface}
\end{eqnarray}

This formula allows us to integrate  $\dot{\alpha}$ at the interval $t\in(t_i,t_{i+1})$ and we can  write the parameters of the system at  $t=t_{i+1}$:
\begin{eqnarray*}
\alpha_{i+1}=F_1(\alpha_i,\dot{\alpha}_i,u_i,dt),
\\
\dot{\alpha}_{i+1}=F_2(\alpha_i,\dot{\alpha}_i,u_i,dt),
\\
A_{i+1}=F_3(\alpha_i,\dot{\alpha}_i,u_i,dt)
\end{eqnarray*}

One can obtain the first integral for WIP on the hard horizontal surface (\ref{EqPendulumOnWheel0WithStab}) if one assumes  $\epsilon=0$.

The mathematical model for the WIP on the soft surface does not integrate obviously. Nevertheless this model can be written in the form:
\begin{eqnarray*}
\dot{A}=\alpha,\quad
\dot{\alpha}=a,\\
\dot{a}=\sin(\alpha)-(\cos(\alpha-z)\dot{b}+\sin(\alpha-z)b^2)\rho-2\frac{\rho}{\zeta}u,\\
\dot{\beta}=b,\quad
\dot{b}\in\frac{1}{(2+\zeta)\rho}F(\alpha,a,\dot{a},b).
\end{eqnarray*}

One can obtain the numeric solution of this differential inclusion at the interval $t\in(t_i,t_{i+1})$. Let us define the map: 
$$
(A_n,\alpha_n,a_n,\beta_n,b_n)
\to
(A_{n+1},\alpha_{n+1},a_{n+1},\beta_{n+1},b_{n+1}).
$$ Formally this map can be written like a discrete dynamical system:
\begin{eqnarray*}
\mathbf{X}_{n+1}=\mathbf{F}(\mathbf{X}_n),\quad \text{where}\quad  \mathbf{X}_n=(A_n,\alpha_n,a_n,\beta_n,b_n,u_n).
\end{eqnarray*}

\section{Observability of the mathematical model for WIP}
\label{secObservation}

In this section we consider the set of the data necessary for the observability of the parameters of the mathematical model for WIP  (\ref{inclusionForPendulumOnWheelFullWithStabAndDump}).

The angle of the tilt for the pendulum is defined by gyroscope. The gyroscope can be work in two different cases. The first one it defines the angle of the pendulum and the second one it define the angle velocity for the pendulum. In the second case one should integrates the angle velocity to obtain the pendulum angle. Below we will use the gyroscope in the mode of angle velocity. This means the value of the angle velocity  $\dot{\alpha}$ is known at the moment of the measurement. 

Besides the gyroscope we assume as existing the tilt sensor. This sensor define the linear acceleration of the pendulum in the plane of the moving of WIP. 

Let us define the coordinates as  $(x,y)$, where $x$ is the horizontal coordinate and $y$ is the vertical one. The projections of the acceleration vector on the coordinate axes $Ox$ and $Oy$ one can write as follows:
\begin{eqnarray*}
\ddot{x}&=&\ddot{\beta}r\cos(z(\beta))+
\ddot{\alpha}l\sin(\alpha),
\\
\ddot{y}&=&\ddot{\beta}r\sin(z(\beta))+
\ddot{\alpha}l\cos(\alpha)+g.
\end{eqnarray*}
It is convenient to write these formulas in the form:
\begin{eqnarray*}
\ddot{\alpha}l\cos(\alpha+z)&=&-\ddot{x}\sin(z)+
\ddot{y}\cos(z)-g\sin(z),
\\
\ddot{\beta}r\cos(\alpha+z)&=&\ddot{x}\cos(\alpha)-
\ddot{y}\sin(\alpha)+g\sin(\alpha).
\end{eqnarray*}

One more sensor is the encoder. This sensor allows us to define the turn of the wheel. The data from the encoder allow to obtain the mean value of the angle velocity of the wheel as value of the difference between the current value of the turn angle of the wheel and another one value at previous measurement:
$$
\dot{\beta}\sim \frac{\beta(t)- \beta(t-\Delta t)}{\Delta t}.
$$

Let us consider the system for WIP on the horizontal surface (i.e. $z=0$):
$$
\ddot{\alpha}l\cos(\alpha)=\ddot{y}
$$
$$
\ddot{\beta}r\cos(\alpha)=\ddot{x}\cos(\alpha)-
\ddot{y}\sin(\alpha)+g\sin(\alpha).
$$ 
The value $a_1=\dot{\alpha}$ is known from the sensor. Let us define by $a_2(t)=\ddot{y}/l$, $b_2=\ddot{x}/r$, $\gamma=g/l$  and $b_1=\dot{\beta}$. Then the dynamical system  (\ref{inclusionForPendulumOnWheelFullWithStabAndDump})  can be written as the system of the trigonometric equation and the inclusion:
\begin{eqnarray}
\frac{\mathit{a_2}}{\cos{\left( \alpha \right) }}=-\left( \cos{\left( \alpha -z\right) } \left( \frac{\sin{\left( \alpha \right) } \gamma }{\cos{\left( \alpha \right) } \rho }-\frac{\mathit{a_2} \sin{\left( \alpha \right) }}{\cos{\left( \alpha \right) } \rho }+\mathit{b_2}\right) +{\mathit{b_1^2}} \sin{\left( \alpha -z\right) }\right)  \rho 
\nonumber
\\
-\frac{2 u \rho }{\zeta }+\sin{\left( \alpha \right) },
\label{EqForObservationAlphaU}
\\
\left( \zeta +2\right) \, \left( \frac{\sin{\left( \alpha \right) } \gamma }{\cos{\left( \alpha \right) } \rho }-\frac{\mathit{a_2} \sin{\left( \alpha \right) }}{\cos{\left( \alpha \right) } \rho }+\mathit{b_2}\right)  \rho 
\in
\left\{
\begin{array}{cc}
f-\nu\sgn(b_1),\quad b_1\not=0;
\\
(-\nu,\nu),\quad b_1=0 \cup |f|<\nu;
\\
f,\quad b_1=0 \cup |f|\ge\nu;
\end{array}
\right.
\label{InclusionForObservationAlphaU}
\end{eqnarray}
where
$$
f=
\frac{2 u}{\rho } -
\left( \frac{\mathit{a_2} \cos{\left( \alpha -z\right) }}{\cos{\left( \alpha \right) }}-{\mathit{a_1^2}} \sin{\left( \alpha -z\right) }\right)  \zeta -\sin{(z)}.
$$
The angle of the pendulum $\alpha$ and the control torque 
$u$ are the unknown variables in the system  (\ref{EqForObservationAlphaU}), (\ref{InclusionForObservationAlphaU}).

If $b_1\not=0$ or $b_1=0 \cup |f|\ge\nu$ then the inclusion (\ref{InclusionForObservationAlphaU}) turn to the following equation: 
\begin{eqnarray*}
\left( \zeta +2\right) \, \left( \frac{\sin{\left( \alpha \right) } \gamma }{\cos{\left( \alpha \right) } \rho }-\frac{\mathit{a_2} \sin{\left( \alpha \right) }}{\cos{\left( \alpha \right) } \rho }+\mathit{b_2}\right)  \rho 
=
\\
\frac{2 u}{\rho } -
\left( \frac{\mathit{a_2} \cos{\left( \alpha -z\right) }}{\cos{\left( \alpha \right) }}-{\mathit{a_1^2}} \sin{\left( \alpha -z\right) }\right)  \zeta -\sin{(z)}-\nu\sgn(b_1).
\end{eqnarray*}

As a result one get the system of the equations for  $\alpha,u$. The control torque $u$ can be easy found through the trigonometric functions of  $\alpha$ and hence one get the trigonometric equation for $\alpha$ . 

As $\{b_1=0\} \cup \{|f|<\nu\}$ the angle of the pendulum should be solution of the inequality:
$$
\frac{-\nu}{\zeta+2}-b_2\rho< (\gamma -\mathit{a_2})\tan(\alpha) <\frac{\nu}{\zeta+2}-b_2\rho.
$$
Here one get the observed parameter  $u$. To obtain the integral term $A$ of the PID controller one should use the following formula:
$$
A=u-\frac{k_1}{k_3}\alpha-\frac{k_2}{k_3}a_1.
$$

\begin{theorem}
Let one know the values of the acceleration   $(\ddot{x},\ddot{y})$, angle velocity of the pendulum  $\dot{\alpha}$ and angle velocity of the wheel $\dot{\beta}$, then the observed dynamical system is solution of the trigonometric equation  (\ref{EqForObservationAlphaU}) and the inclusion (\ref{InclusionForObservationAlphaU}).
\end{theorem}
The equations for small values of  $\nu,\alpha,\ddot{y},\dot{\alpha},\ddot{x},\dot{\beta}$ and ($z\equiv0$) can be written in the following form:
\begin{eqnarray}
\mathit{a_2} 
\sim
-\frac{2 \rho  u}{\zeta }+\left(1-\gamma\right)  \alpha -\rho \, \mathit{b_2}, 
\nonumber\\
(\zeta +2)(  \rho \, b_2+  \gamma  \alpha)
\in
\left\{
\begin{array}{cc}
\sim\frac{2 u}{\rho } -\zeta a_2-\sgn(b_1)\nu, \quad b_1\not=0;
\\
\sim(-\nu,\nu),\quad  b_1=0 \cup |-\zeta a_2+2u/\rho|<\nu;
\\
\sim\frac{2 u}{\rho } -\zeta a_2,\quad b_1=0\cup|-\zeta a_2+2u/\rho|\ge\nu.
\end{array}
\right.
\label{InclusionForSmallParameters}
\end{eqnarray}

\begin{corollary}
The important case for the WIP on the hard horizontal is more simplest.  In particular the angle $\alpha$ is the solution of the equation:
\begin{eqnarray}
\rho^2\sin{\left( \alpha \right) } \left( 2 \gamma  \zeta -2 \mathit{a_2} \zeta +4 \gamma -4 \mathit{a_2}\right) +
\nonumber\\
\zeta\sin{\left( 2 \alpha \right) } \left( -{\mathit{a_1^2}} {{\rho }^{2}}+{{\mathit{b_1}}^{2}} \rho +\gamma -\mathit{a_2}-1 \right) =
\\
\rho^2\cos{\left( \alpha \right) } \left( -2 \mathit{b_2} \zeta \, \rho -4 \mathit{b_2}\,\rho-2 \mathit{a_2} \zeta \right) -
\nonumber\\
\mathit{b_2} \cos{\left( 2 \alpha \right) } \zeta  \rho 
-\mathit{b_2} \zeta  \rho -2 \mathit{a_2}\, \zeta 
\nonumber
\label{EqForObservedAlphaOnSolidHorizontal}
\end{eqnarray}
\end{corollary}

For small values of  $\nu,\alpha,\ddot{y},\dot{\alpha},\ddot{x},\dot{\beta}$ we get:

$$
\alpha\sim -\frac{\left( {{\rho }^{2}}+1\right)  \zeta \, \mathit{a_2}+\left( \left( {{\rho }^{3}}+\rho \right)  \zeta +2 {{\rho }^{3}}\right) \, \mathit{b_2}}{\left( \left( {{\rho }^{2}}+1\right)  \gamma -1\right)  \zeta +2 {{\rho }^{2}} \gamma },
$$
$$
u\sim
-\frac{\left( \rho \, {{\zeta }^{2}}+2 \rho  \gamma  \zeta \right) \, \mathit{a_2}+\left( {{\rho }^{2}}\, {{\zeta }^{2}}+2 {{\rho }^{2}} \zeta \right) \, \mathit{b_2}}{\left( \left( 2 {{\rho }^{2}}+2\right)  \gamma -2\right)  \zeta +4 {{\rho }^{2}} \gamma }.
$$

\section{The observability and the stochastic properties}
\label{secStochasticPropOfObservedData}

\begin{figure}
\includegraphics[scale=0.45]{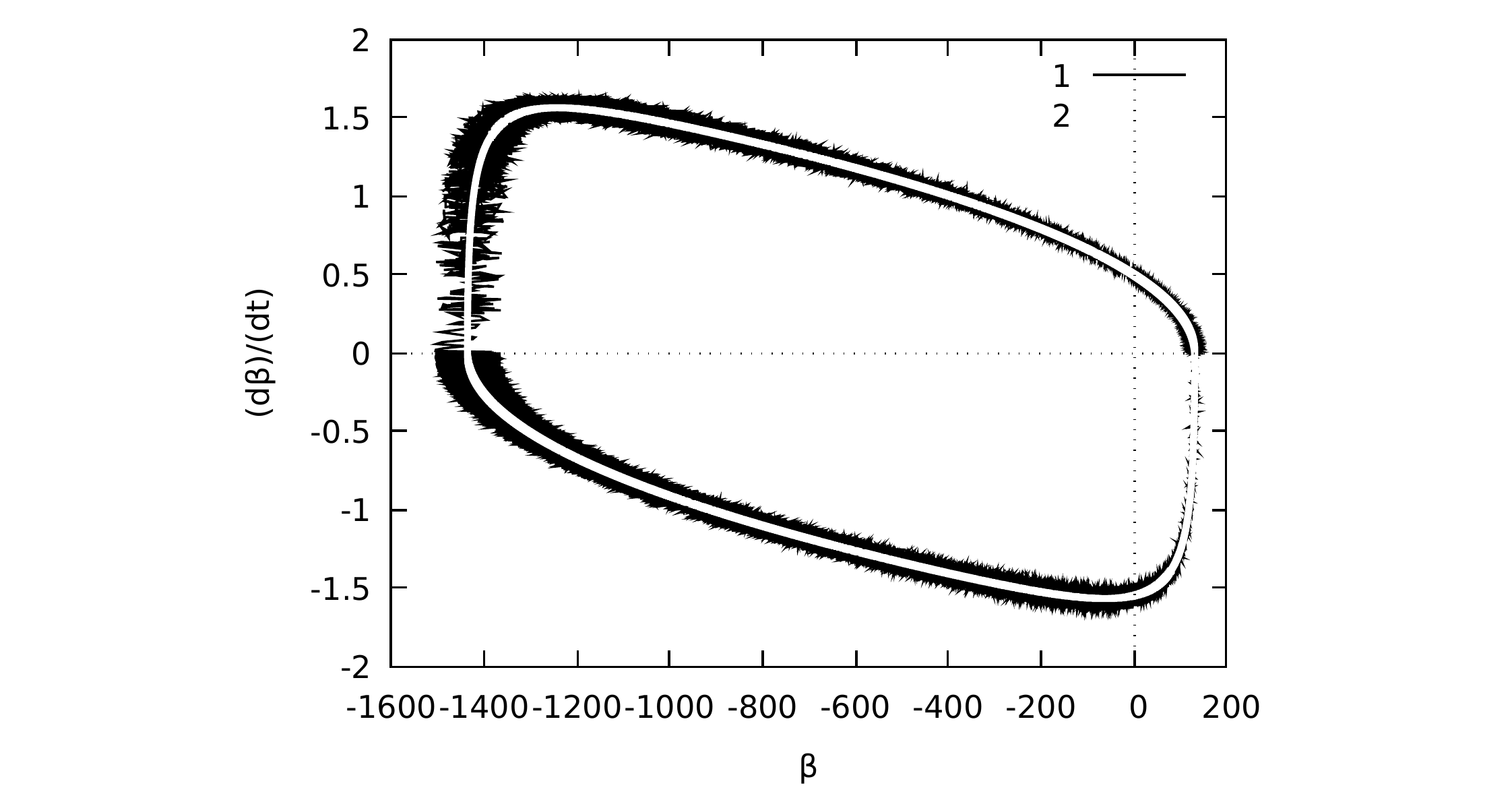}
\caption{
The observed phase curve on the plane  $(\beta,\dot{\beta})$ are black. This curve is obtained using the data from the sensors as $z\equiv0$. The white line is the result of numeric solution for the system for WIP on soft surface. The   feedback controller use the tilt sensor and the gyroscope (\ref{inclusionForPendulumOnWheelFullWithStabAndDump}). The parameters of the system are following: $\rho=0.2,\zeta=10,\, \nu=0.05,\gamma=1$, the PID coefficients are: $k_1=1.7, k_2=0.2, k_3=0.02$. The relative errors are uniform distributed data at the interval $(-0.02,0.02)$. The dynamic system 
(\ref{inclusionForPendulumOnWheelFullWithStabAndDump}) solved at $A\sim 0.2385,\,\alpha=0.02$, $\dot{\alpha}=0,\,\beta=0,\dot{\beta}=0.5$ by Runge-Kutta method of fourth-order method with the step $0.1$. 
}
\label{figSolutionAndObservation}
\end{figure}

The value of the controlling torque at $t_i$ is defined by the measurement of observed values of the parameters of the dynamic system.  The obtained data from the sensors  and the computed observed data at  $t_i$ will be denoted by variables with upper symbol $\breve{\bullet}$.

The absolute errors and the dispersion are known for typical sensors. Below we will assume that we know the standard deviation $\sigma_j$ of the measured data at $t=t_j$.

Let the measured data be following:
\begin{eqnarray*}
\breve{a}_1=\dot{\alpha}+\delta^{(1)},\, 
\breve{a}_2=\ddot{y}/l+\delta^{(2)},\,
\breve{b}=\beta+\delta^{(3)},\,
\breve{b}_1=\dot{\beta}+\delta^{(4)},\,
\breve{b}_2=\ddot{x}/r+\delta^{(5)}.
\end{eqnarray*}
Here $\delta^{(i)}$ is stochastic error. 

Let us consider the WIP on the hard horizontal ($\nu=0$ and $z\equiv0$). We will assume the errors are small and one can use a linear system for find observable values of $\alpha,u$.
\begin{eqnarray*}
\frac{\mathit{a_2}}{\cos{\left( \alpha \right) }}=-\left( \cos{\left( \alpha \right) } \left( \frac{\sin{\left( \alpha \right) } \gamma }{\cos{\left( \alpha \right) } \rho }-\frac{\mathit{a_2} \sin{\left( \alpha \right) }}{\cos{\left( \alpha \right) } \rho }+\mathit{b_2}\right) +{\mathit{b_1^2}} \sin{\left( \alpha \right) }\right)  \rho -
\\
\frac{2 u \rho }{\zeta }+\sin{\left( \alpha \right) }
\\
\left( \zeta +2\right) \, \left( \frac{\sin{\left( \alpha \right) } \gamma }{\cos{\left( \alpha \right) } \rho }-\frac{\mathit{a_2} \sin{\left( \alpha \right) }}{\cos{\left( \alpha \right) } \rho }+\mathit{b_2}\right)  \rho =\frac{2 u}{\rho } -\left( \mathit{a_2}-{\mathit{a_1^2}} \sin{\left( \alpha \right) }\right)  \zeta
\end{eqnarray*}

One can derive the equation for $\alpha$:
\begin{eqnarray}
&
\{((-\gamma +a_1^2\cos(\alpha)+a_2)\zeta -2\gamma+2a_2)\rho^2 - 
\nonumber
\\
&
b_1^2\cos(\alpha)\zeta\rho + a_2\cos(\alpha) \zeta \}
\sin(\alpha)=
\nonumber
\\
&
(b_2 \cos(\alpha) \zeta +2 b_2\cos(\alpha))\rho^3+ (\sgn(b_1)\cos(\alpha)\nu +a_2\cos(\alpha)\zeta) \rho^2 
+
\nonumber
\\
&
b_2\cos(\alpha)^2 \zeta\rho +a_2\zeta.
\end{eqnarray}
 
For small $\ddot{y},\alpha,\dot{\alpha},\ddot{x},\dot{\beta},\nu$ we obtain the formulas for errors of the observed data $\alpha$ and $A$:
$$
\breve{\alpha}\sim\alpha-
\frac{((\zeta+2)\rho^2+\zeta)\rho}
{(\zeta+2)\gamma\rho^2+(\gamma-1)\zeta}
\delta^{(5)}+
\frac{(\rho^2 +1)\zeta}{(\zeta+2)\gamma\rho^2+(\gamma-1)\zeta}
\delta^{(2)}
$$

\begin{eqnarray}
\breve{u}\sim u+
\frac{(\gamma -1)\zeta\nu\rho }{\left( 2 \gamma  \zeta +4 \gamma \right) \, {{\rho }^{2}}+\left( 2 \gamma -2\right)  \zeta }
(\sgn(b_1+\delta_4)-\sgn(b_1))
\nonumber
\\
-
\frac{(\zeta^2+2\gamma \zeta )\rho }{\left( 2 \gamma  \zeta +4 \gamma \right) \, {{\rho }^{2}}+\left( 2 \gamma -2\right)  \zeta }\delta^{(2)}
-\frac{(\zeta^2+2\zeta)\rho^2}{\left( 2 \gamma  \zeta +4 \gamma \right) \, {{\rho }^{2}}+\left( 2 \gamma -2\right)  \zeta }\delta^{(5)},
\nonumber
\end{eqnarray}
hence:
$$
\breve{A}\sim \breve{u}-\frac{k_1}{k_3}\breve{\alpha}-
\frac{k_2}{k_3}(\dot{\alpha}+\delta^{(1)}).
$$

{\bf Remark.}While the WIP moves on the soft horizontal the stochastic layer appears near the hyperplane $\dot{\beta}=0$. The width of this layer is  $\min\{\delta^{(4)}\}\le\dot{\beta}\le\max\{\delta^{(4)}\}$. In this layer the stochastic error can be  $\pm\nu$ when $|f|>\nu$. It is important the value of this errors defines by value of the rolling resistance for the wheel and does not depend on the error of the encoder.

\begin{figure}
\includegraphics[scale=0.5]{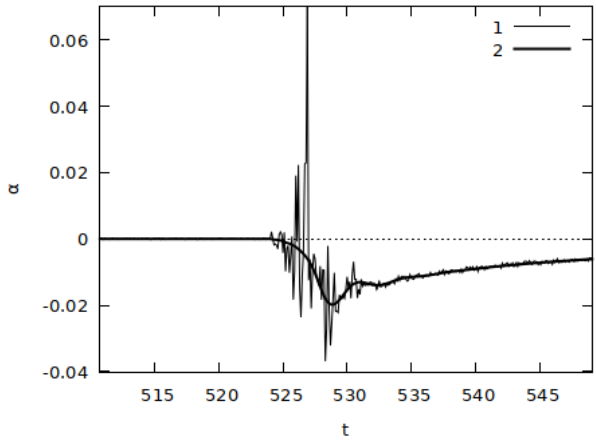}
\caption{
On the picture one can see the result of numeric modelling for the angle of the pendulum of WIP on the soft horizontal. The horizontal axis defines values of the time and the vertical axis defines the value of the angle of the pendulum. The curve 2 is the numeric value and the curve 1 shows the model of the observed data. The observed data  $\breve{\ddot{\alpha}}_i$ and $\breve{\dot{\alpha}}_i$ are modelled using current values of $\ddot{\alpha}_i$ and $\dot{\alpha}_i$ with the uniform distribution of the relative error at the interval $(-0.05,0.05)$. The value of the angle  $\breve{\alpha}_i$ is defined as the observed calculated using the angle acceleration and angle velocity from the equation  (\ref{EqForObservedAlphaOnSolidHorizontal}). The values  $\breve{\alpha}$ and $\breve{\dot{\alpha}}$ were used to obtain  $\breve{A}$ by integrating by the trapezoidal rule. The  value of the control torque   $u_{i+1}$ was obtained at the interval $t\in(t_i,t_{i+1})$. The system of the equations was solved at the interval $t\in(t_i,t_{i+1})$ with the constant value of the control torque  $u=u_{i+1}$ by Runge-Kutta fourth-order method with the step equals by $0.1$.
}
\label{figAlphaPIDSolidHorizontalFiltered}
\end{figure}

The data with stochastic errors are used in the PID controller. As a result the stochastic perturbations appear in the mathematical model for the WIP  (\ref{EqPendulumOnWheel0WithStab} and in the systems  (\ref{EqPendulumOnWheel1WithStab}) and (\ref{inclusionForPendulumOnWheelFullWithStabAndDump}). Therefore the mathematical model with the PID controller looks like the stochastic differential inclusion  (\ref{inclusionForPendulumOnWheelFullWithStabAndDump}).
In partial the results for the observed values of  $(\beta,\dot{\beta})$ and $\alpha$ with stochastic errors are showed in the figures  \ref{figSolutionAndObservation} and  \ref{figAlphaPIDSolidHorizontalFiltered}.

{\bf The remark about filtering data}

The current values of the parameters of the dynamic system for WIP one can obtain by the different approaches. 

The first one is the integration of the differential inclusion as the predetermined process. Such approach gives the errors at any step of the integration because of two causes. First of all this errors appear because of the errors in the initial data on the first step of integration. One more cause of the appearance of the errors is the inaccuracy of the mathematical model.

Another one approach is to use the observability of this system. This case does not needed to integrate the differential inclusion. But the errors appears in the current moment because of the errors of the measurement of the data using the sensors.

To minimize the quadratic deviation of the data one can combine the observed data and the forecasted data using the deterministic methematical model. Such algorithms are called as the filters. The filters for the linare system are  well-knowing, see  \cite{Kalman1960}, \cite{KalmanBucy1961}, \cite{BrammerSiffling1975D}. FOr nonlinear smooth systems like the WIP on the hard surface is convenient the generalized Kalman's filter, see \cite{AthansWishnerBertolini1968}, \cite{AustinLeondes1981}. But for the considered here case of WIP on the soft surface the generalized Kalman's filter is  not appropriated because of non-lineared the dinamical system in the neighbouhood of the hypersurface  $\dot{\beta}=0$. One of the opportunity to use  filtering in such case is the sigma-point filter, see \cite{JulierUhlrnannDurrant-Whyte1995}, \cite{JulierUhlmann1997}.

\section{Stochastic properties for WIP on soft horizontal}
\label{secStochasticPropOfPendulum}

\begin{figure}
\hspace{-1.5cm}
\includegraphics[scale=0.4]{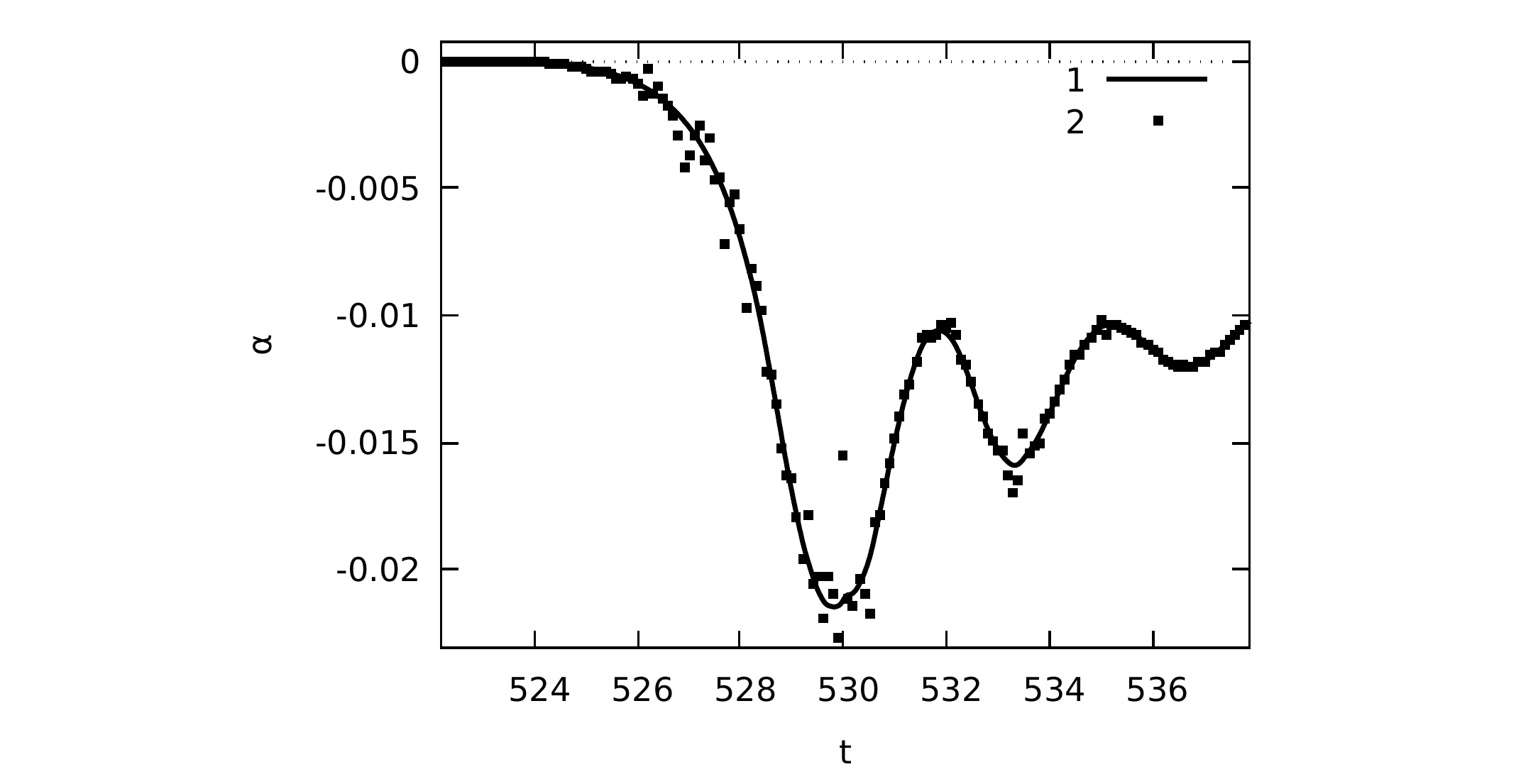}
\hspace{-1.5cm}
\includegraphics[scale=0.4]{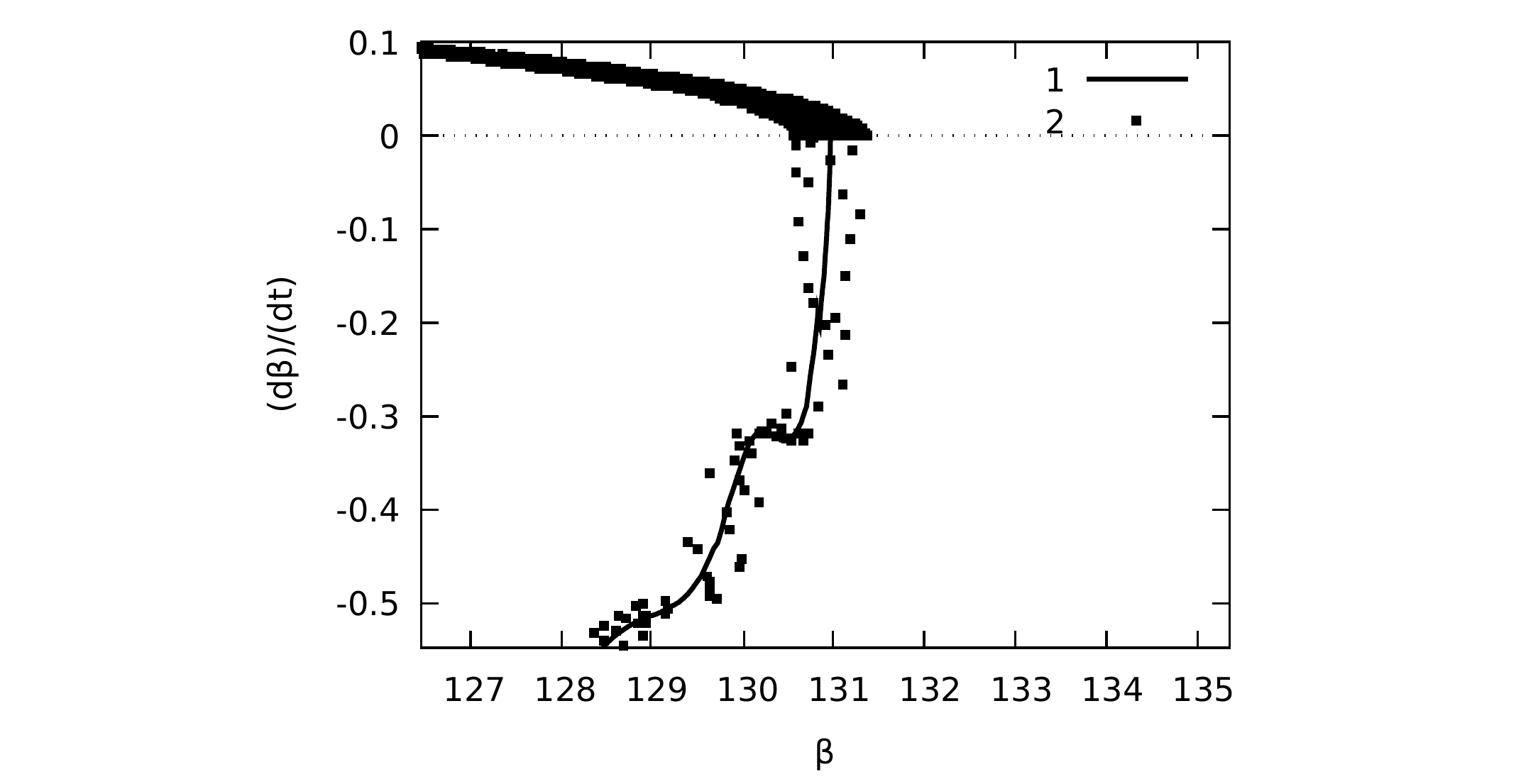}
\caption{
In this picture one can see the result of the numeric modelling for the behaviour of the angle for the pendulum at $\xi=10$, $\rho=0.2$, $\nu=0.05$, $\gamma=1$, $k_1=1.7$, $k_2=0.2$, $k_3=0.02$. On the left picture the horizontal axis shows the time variable $t$ and the vertical axis shows the angle of the pendulum $\alpha$. On the right picture the horizontal axis shows $\beta$ and the vertical axis shows $\dot{\beta}$. The step of the change of the    control torque is $0.1$. The line is the solution under the discrete control. The results of the measurements  $\breve{\ddot{\alpha}}_i$, $\breve{\dot{\alpha}}_i$ and $\breve{\dot{\beta}}_i$ are modelled by the current values $\ddot{\alpha}_i$, $\dot{\alpha}_i$ and $\dot{\beta}_i$ with the uniform distribution of the  relative errors at the interval $(-0.003,0.003)$. The value of the angle $\breve{\alpha}_i$ is defined as the observed data through the angle acceleration and the angle velocity using (\ref{inclusionForPendulumOnWheelFullWithStabAndDump}). The value $\breve{A}$ is computed using $\breve{\alpha}$, $\breve{\dot{\alpha}}$ integrating by the trapezoidal method. It allows to obtain the control torque $u_{i+1}$ at the interval $t\in(t_i,t_{i+1})$, where $t_{i+1}-t_i=dt$. At $t\in(t_i,t_{i+1})$ the system for the WIP on the soft horizontal is solved for the constant value the control torque $u=u_{i+1}$ by the Runge-Kutta method of the fourth order with the step equals $0.01$.
}
\label{figBetaDBetaPIDStochastic}
\end{figure}

Let the interval $dt$ between the moments of the measurements be small. Then one can see at the dynamic system as a determined dynamic system  (\ref{inclusionForPendulumOnWheelFullWithStabAndDump}) with stochastic perturbation. The stochastic perturbation is contained in the control torque:
$$
\tilde{u}_i= k_1\tilde{\alpha}_i +k_2\tilde{\dot{\alpha}} +k_0\tilde{A}_i.
$$

In the work \cite{Kiselev2020} (see also preprint \cite{Kiselev2020arxiv1}) it was shown that the unperturbed dynamical system with the PID controller has the attractor as $\sgn(\dot{\beta})=\pm1$. This attractor is a line belonged the fifth-dimensional phase space:  
$(A,\alpha,\dot{\alpha},\beta,\dot{\beta})=(A_{\pm},0,0,0,\dot{\beta})$, where 
$$
\frac{2}{\rho}k_0A_{\pm}=\pm\nu.
$$

On this line the system for WIP is unstable and due to the perturbations crosses to the trajectory with changing of the signum of rotation of the wheel from  $\sgn(\dot{\beta})=\pm1$ to $\sgn(\dot{\beta})=\mp1$. As a result the numeric modelling gives the trajectory like the hysteresis loop, see \cite{Kiselev2020}.

Let us consider here the impact of the stochastic perturbation on the stability for the hysteresis loop. The typical trajectory for the system with the stochastic perturbation is shown on the figure \ref{figBetaDBetaPIDStochastic}.

\begin{theorem}
The line $(A_{\pm},0,0,0,\dot{\beta})$ as $\sgn(\dot{\beta})=\pm1$ is the attractor for the stochastic system (\ref{inclusionForPendulumOnWheelFullWithStabAndDump}).
\end{theorem}

This theorem is corollary from the results of  \cite{Kiselev2020} concerning the stability of the line $(A_{\pm},0,0,0,\dot{\beta})$ as $\sgn(\dot{\beta})=\pm1$ for pure determined dynamical system for WIP under the PID controller and the theorem about stability under constantly perturbations \cite{Krasovskii1959Eng}.

The layer $|\dot{\beta}|\le \max\{\delta^{(3)}\}$ appears in the stochastic system near the hyperplane  $\dot{\beta}=0$. In this layer the term $\nu\sgn(\dot{\beta})$ takes the random values  $\pm\nu$ at  $t\in(t_i,t_{i+1})$. 

There exists the small neighbourhood ($\Delta_{\pm}$) near the unstable lines $(A_{\pm}$, $0$, $0$, $0$, $\dot{\beta})$, where  can be obtained four typical cases:

\begin{itemize}
\item Let $\dot{\beta}>0$, $A<A_{+}$ 
\begin{itemize}
\item and $\sgn{\breve{\dot{\beta}}}=1$, then the trajectory is kept in the neighbourhood of the line $(A_{+},0,0,0,\dot{\beta})$;
\item  and $\sgn{\tilde{\dot{\beta}}}=-1$, then the trajectory is kept in the neighbourhood of the line $(A_{+},0,0,0,\dot{\beta})$.
\end{itemize}
\item Let $\dot{\beta}<0$, $A>A_{-}$ 
\begin{itemize}
\item and $\sgn{\breve{\dot{\beta}}}=-1$, then the trajectory is kept in the neighbourhood of the line $(A_{-},0,0,0,\dot{\beta})$;
\item  and $\sgn{\breve{\dot{\beta}}}=1$, then the trajectory is kept in the neighbourhood of the line $(A_{-},0,0,0,\dot{\beta})$.
\end{itemize}
\end{itemize}

The sequence of the changes of the trajectories at the neighbourhoods of the lines $(A_{\pm},0,0,0)$ leads to the appearance of the hysteresis loop at the phase plane  $(\beta,\dot{\beta})$, see figure \ref{figSolutionAndObservation}. 

Here it is important for applications the average time,which the WIP spends in the neighbourhood of the upper position.

The time between the sequence measurements is equal  $dt$. Let the trajectory be in the neighbourhood  $\Delta_{\pm}$ of the unstable line.  The probability of   $\sgn(\delta^{(3)})=\pm1$ in primary order as $\Delta_\pm\to0$ equals   $p_{\pm}\sim1/2$.  The average time for trajectory in this neighbourhood is following: 
$$
T_0=dt\sum_{n=1}^\infty\frac{n}{2^{n}}=2dt.
$$

\begin{theorem}
The average time spending at $\Delta$-neighbourhood of the unstable lines $(A_{\pm},0,0,0,\dot{\beta})$ for the stochastic system  (\ref{inclusionForPendulumOnWheelFullWithStabAndDump}) equals  $2dt$, where $dt$ is the time between the sequenced measurements of the state for the system.
\end{theorem}

\section{Conclusion}
The system for the WIP with discrete control by the PID controller is stochastic due to the errors of the measurements. The stabilising of the WIP on the soft surface leads to the appearance of the hysteresis loop in the plane of the phase variables $\beta,\dot{\beta}$. The average time spending near the upper position was calculated.


\begin{thebibliography}{10}

\bibitem{NasirRajaIsmailAhmad2010}
M.~A.~Ahmad A.~N. K.~Nasir, R. M. T. Raja~Ismail.
\newblock Performance comparison between sliding mode control (smc) and pd-pid
  controllers for a nonlinear inverted pendulum system.
\newblock {\em World Academy of Science, Engineering and Technology},
  71:122--127, 2010.

\bibitem{AstromHagglund1994}
K.J. \AA{}str\"{o}m and T.~H\aa{}gglund.
\newblock {\em PID controllers,2dn edition}.
\newblock 1994.

\bibitem{AustinLeondes1981}
J.~W. Austin and C.~T. Leondes.
\newblock Statistically linearized estimation of reentry trajectories.
\newblock 17:54--61.

\bibitem{BrammerSiffling1975D}
K.~Brammer and G.~Siffling.
\newblock {\em Kalman‐Bucy‐Filter, Deterministische Beobachtung und
  stochastische Filterung}.
\newblock Methoden der Regelungstechnik.

\bibitem{Formalskii2016Eng}
A.M. Formalskii.
\newblock {\em Stabilisation and Motion Control of Unstable Objects. Series:De
  Gruyter Studies in Mathematical Physics 33}.
\newblock 2016.

\bibitem{MMA7361L}
Freescale Semiconductor.
\newblock {\em $\pm1.5g, \pm6g$ Three Axis Low-g Micromachined Accelerometer},
  04 2008.
\newblock Rev. 0.

\bibitem{mpu6000}
InventSense.
\newblock {\em MPU-6000/MPU-6050 Product Specification}, 08 2013.
\newblock Rev. 3.4.

\bibitem{JulierUhlmann1997}
S.J. Julier and J.K. Uhlmann.
\newblock A new extension of the kalman filter to nonlinear systems.
\newblock page 182–193.

\bibitem{KalmanBucy1961}
Bucy~R.S Kalman, R.E.
\newblock New results in linear filtering and prediction theory.
\newblock 83:95–108.

\bibitem{Kalman1960}
R.~E. Kalman.
\newblock A new approach to linear filteringand prediction problems.
\newblock 82(D):35--45.

\bibitem{Kiselev2020}
O.M. Kiselev.
\newblock Stabilization of the wheeled inverted pendulum on a soft surface.
\newblock {\em Russian Journal of Nonlinear Mechanics}, 16(3).

\bibitem{Kiselev2020arxiv1}
O.M. Kiselev.
\newblock Stabilization of the wheeled inverted pendulum on a soft surface.
  arxiv:2006.05450.

\bibitem{Krasovskii1959Eng}
N.N. Krasovskii.
\newblock {\em Nekotorye zadachi teorii ustoichivisti dvizheniya}.
\newblock FizMatLit.

\bibitem{AthansWishnerBertolini1968}
A.~Bertolini M.~Athans, R. P.~Wishner.
\newblock Suboptimal state estimation for continuous-time nonlinear systems
  from discrete noisy measurements.
\newblock 13:504--518.

\bibitem{MartynenkoFormalskii2013Eng}
Yu.~G. Martynenko and A.~M. Formal’skii.
\newblock Controlled pendulum on a movable base.
\newblock {\em Mechanics of Solids}, 48:6--18, 2013.

\bibitem{HatakeyamaShimada2008}
A.~Shimada N.~Hatakeyama.
\newblock Movement control using zero dynamics of two-wheeled inverted pendulum
  robot.
\newblock {\em 10th IEEE international workshop on advanced motion control},
  pages 38--43, 2008.

\bibitem{SamiMichalskaAngeles2007}
D.~S. Nasrallah, H.~Michalska, and J.~Angeles.
\newblock Controllability and posture control of a wheeled pendulum moving on
  an inclined plane.
\newblock {\em IEEE TRANSACTIONS ON ROBOTICS}, 23(3):564--577, 2007.

\bibitem{PathakFranchAgrawal2005}
Kaustubh Pathak, Jaume Franch, and Sunil~K. Agrawal.
\newblock Velocity and position control of a wheeled inverted pendulum by
  partial feedback linearization.
\newblock {\em IEEE TRANSACTIONS ON ROBOTICS}, 21:505--513, 2005.

\bibitem{ChanStolHalkyard2013}
C.~R.~Halkyard R.~P. M.~Chan, K. A.~Stol.
\newblock Review of modelling and control of two-wheeled robots.
\newblock {\em Annual Reviews in Control}, 37:89--103, 2013.

\bibitem{JulierUhlrnannDurrant-Whyte1995}
Jeffrey K.~Uhlrnann Simon J.~Julier and Hugh~F. Durrant-Whyte.
\newblock A new approach for filtering nonlinear systems.
\newblock pages 1628--1632.

\end{thebibliography}
\end{document}